\documentclass[twocolumn,showpacs,preprintnumbers,amsmath,amssymb]{revtex4}
\usepackage{graphicx}% Include figure files
\usepackage{dcolumn}% Align table columns on decimal point
\usepackage{bm}% bold math

\usepackage{bbm}
\usepackage{mathrsfs}

\newtheorem{observation}{Observation}
\newtheorem{lemma}{Lemma}

\renewcommand{\phi}{\varphi}
\renewcommand{\theta}{\vartheta}

\newcommand{\ov}[1]{\overline{#1}}

\newcommand{\be}{\begin{equation}}
\newcommand{\ee}{\end{equation}}
\newcommand{\eea}{\end{eqnarray}}
\newcommand{\bea}{\begin{eqnarray}}

\newcommand{\mean}[1]{\ensuremath{\langle{#1}\rangle}}

\newcommand{\eins}{\ensuremath{\mathbbm 1}}
\newcommand{\qed}{\ensuremath{\hfill \Box}}

\newcommand{\ketbra}[1]{\ensuremath{| #1 \rangle \langle #1 |}}

\newcommand{\ket}[1]{\ensuremath{|#1\rangle}}

\newcommand{\kommentar}[1]{}

\begin{document}

%\preprint{APS/123-QED}

\title{Entanglement verification for quantum key distribution 
systems with an underlying bipartite qubit-mode structure
}
\author{Johannes Rigas$^1$, Otfried G\"uhne$^2$ and Norbert L{\"u}tkenhaus$^1$}
\affiliation{$^1$Quantum Information Theory Group, Institut f\"ur
Theoretische Physik I, and Max-Planck Research Group, Institute
of Optics, Information and Photonics, Universit\"{a}t
Erlangen-N\"{u}rnberg, Staudtstra{\ss}e 7/B2, 91058 Erlangen,
Germany\\
$^2$Institut f\"ur Quantenoptik und Quanteninformation,
\"Osterreichische Akademie der Wissenschaften,
A-6020 Innsbruck, Austria
}

\date{\today}
\begin{abstract} 
We consider entanglement detection for quantum key distribution 
systems that use two signal states and continuous variable
measurements. This problem can be formulated as a separability 
problem in a qubit-mode system. To verify entanglement, we introduce 
an object that combines the covariance matrix of the  mode with the 
density matrix of the qubit. We derive necessary separability 
criteria for this scenario. These criteria can be readily evaluated 
using semidefinite programming and we apply them to the specific 
quantum key distribution protocol.
\end{abstract}

%\keywords{}
\maketitle

\section{Introduction}

Quantum key distribution (QKD) 
protocols typically distinguish 
two phases: In the first phase, a physical apparatus is used to 
establish correlated data between the sender (called Alice) and 
the receiver (called Bob). This data are described by a  joint 
probability distribution. In the second phase the data are processed 
by classical communication via an authenticated public channel 
employing methods such as post-selection, error correction and 
privacy amplification  to distill a secret key 
(for a review see \cite{gisin02a}). 
 
A necessary precondition for the success of Phase II, i.e., 
for obtaining a secret key, is that the correlations in the 
data show signatures from quantum entanglement \cite{curty04a}. 
This means that the data must originate from an effective entangled 
state (effective, since the quantum state is not shared any more). 
Whenever only partial information on the whole bipartite state is 
available from the data, it means that all possible states 
compatible with the measurement outcomes must be entangled.
If there is a separable state consistent with the data, then
the QKD protocol is not secure.

For this, it makes no conceptual difference whether the entangled 
state is first distributed by an untrusted third party Eve before 
Alice and Bob perform the measurements on the state (so-called
\emph{entanglement-based} schemes, EB, see e.g. \cite{ekert91a}) 
or whether Alice  prepares 
an entangled state first,  measures her part before sending the other 
part through the insecure domain under Eve's control to Bob, who 
performs his measurements (so-called \emph{prepare\&measure} schemes, 
PM, see e.g. \cite{bennett84a, bennett92b}).    

The investigation of entanglement in QKD protocols using discrete 
variables, mainly qubits, was considered before for various 
protocols \cite{curty04a,curty04suba}. For the case where 
Alice and Bob both control a continuous variable system, this issue 
has been addressed in Refs.~\cite{grosshans03a,lance05a}. In this 
paper, we study this problem for the case where Alice owns a 
discrete system, namely a qubit, and Bob owns a mode.

\begin{figure}
\begin{center}
\centerline{\resizebox{0.9\hsize}{!}{\includegraphics*{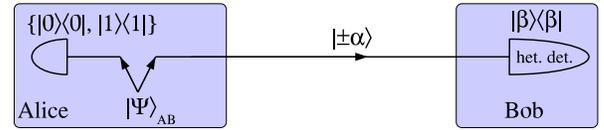}}}
\end{center} 
\caption{\label{fig:Setup} (color online) In the considered QKD scheme Alice effectively 
sends two coherent states $\ket{\pm \alpha}$ to Bob, who performs 
heterodyne measurement, e.g. a projection onto coherent states. Alice's 
state preparation can be thought of as coming from an initial entangled 
state, as described in the text.}
\end{figure}

The protocol we investigate is described as follows (see
Fig.\ref{fig:Setup}): Alice prepares the entangled state 
\begin{equation}
  \label{eq:Prep}
  |\Psi\rangle_{AB} = \sqrt{p_0}\,|0\rangle_A \otimes|\alpha\rangle_B
   + 
   \sqrt{p_1}\,|1\rangle_A\otimes|-\alpha\rangle_B 
\end{equation}
at her site. By projecting the state $|\Psi\rangle\langle\Psi|_{AB}$ 
either onto $|0\rangle\langle0|_A$ or onto $|1\rangle\langle1|_A$, 
Alice effectively sends coherent states   $|\pm \alpha\rangle_B$ 
with \emph{a priori probabilities} $p_0, p_1$ to Bob. The overlap 
of those input states is significantly larger than zero. 
Since Alice keeps her part of the state, her reduced density 
matrix 
\begin{equation}
  \label{eq:rho_A}
  \rho_A := \mbox{Tr}_B(|\Psi\rangle\langle\Psi|_{AB})= 
  \begin{bmatrix}
    p_0 & \sqrt{p_0 p_1}\langle -\alpha|\alpha\rangle \\
   \sqrt{p_0 p_1} \langle\alpha|-\alpha\rangle & p_1
  \end{bmatrix}
\end{equation}
is fixed.

After passing through the insecure domain controlled by Eve, 
Bob receives these states which may have changed, in particular 
affected by loss and noise, and he measures the covariance matrix
of them, for instance by performing heterodyne detection.
The states Bob receives conditioned on which state was sent by
Alice are labeled with $\rho_0$ and $\rho_1$. 
After the measurements, the data are processed by classical
communication in order to obtain a secret key.     

This protocol is similar to the one proposed and implemented
in \cite{lorenz04a} with the difference that in Ref.~\cite{lorenz04a} 
also a strong phase reference, which is necessary for heterodyne 
detection as a local oscillator, has been sent from Alice 
to Bob. Since Eve may also access this phase reference, 
the security analysis for a practical setup is more complicated 
compared with the protocol described above.

The structure of this paper is as follows: With a simplified example,
in Section II we outline that the key idea behind our approach is 
that the outcome states measured by Bob are very pure. In Section III, 
we introduce a description of qubit-mode systems which includes all 
information on the bipartite state accessible with heterodyne detection. 
We note some basic properties of this description and derive a
necessary criterion for separability.
In Section IV these conditions are applied to the special
case of limited knowledge on the whole state in a PM scheme. 
Finally,  a sufficient entanglement criterion is implemented 
numerically where the performance of the criterion is discussed 
with help of an explicit example.

\section{Basic idea behind entanglement detection}

Let us explain the main idea for our entanglement detection
scheme in a simple example. To that aim we will show no 
separable state can be compatible with the data if both 
conditional states $\rho_0,\rho_1$ are pure.

So let us  assume that Bob receives two non-orthogonal, 
non-identical pure states, i.e., 
$\rho_i=|\phi_i\rangle\langle\phi_i|,\,i=0,1$, and
$0<$Tr$(\rho_0\rho_1)<1$. 
In this case, we  can describe the whole bipartite state 
$\rho_{AB}$ as   
\begin{equation}
  \label{eq:rho_AB}
  \rho_{AB} = 
  \begin{bmatrix}
    p_0 \rho_0 & \vline & C^\dag\\ \hline 
    C & \vline  & p_1 \rho_1
  \end{bmatrix},
\end{equation}
with the pure states $\rho_i$ completely known 
due to the tomographical completeness of Bob's 
 heterodyne measurement and two arbitrary matrices 
$C,C^\dag$ of which only the trace is known since
Tr$(C)=(\rho_A)_{10}=\sqrt{p_0 p_1} \langle\alpha|
-\alpha \rangle$.  

The two pure states $\ket{\phi_i}$ span Bob's Hilbert 
space, so we can write down the matrix blocks in 
Eq.(\ref{eq:rho_AB}) in the eigenbasis of $\rho_0$: 
\begin{equation}
  \label{eq:rho_0_AB}
  \rho_{AB} = 
  \begin{bmatrix}
    \begin{array}{cc}
p_0 & 0\\
0 & 0
    \end{array}& \vline&
    \begin{array}{cc}
c_{00}^* & c_{10}^*\\
c_{01}^* & c_{11}^*
    \end{array}
\\ \hline
    \begin{array}{cc}
c_{00} & c_{01}\\
c_{10} & c_{11}
    \end{array}&\vline& p_1 \rho_1
  \end{bmatrix}\,.
\end{equation}
In this representation we can easily implement the 
constraint that $\rho_{AB}$ is positive. Namely, this 
implies that $|c_{01}|=|c_{11}|=0,$ since the element 
$(\rho_{AB})_{22}$ on the diagonal is zero \cite{hornjohnson85a}. 

Under the assumption of separability of $\rho_{AB}$, 
also its partial transpose must be positive 
\cite{peres96a,horodecki96a}. Performing the partial 
transposition leads to the conclusion, that for PPT 
states also $|c_{10}|=0$ has to hold.

For separable states we have therefore in the eigenbasis  
of $\rho_0$
\begin{equation}
  \label{eq:rho_0_AB_cleaned}
  \rho_{AB} = 
  \begin{bmatrix}
    \begin{array}{cc}
p_0 & 0\\
0 & 0
    \end{array}&\vline &
    \begin{array}{cc}
S &  0\\
0 &  0
    \end{array}
\\ \hline
    \begin{array}{cc}
S &  0\\
0 &  0
    \end{array}& \vline &p_1 \rho_1
  \end{bmatrix},
\end{equation}
with $S$ abbreviating Tr$(C)=\sqrt{p_0 p_1}\langle
\alpha|-\alpha\rangle=\sqrt{p_0 p_1}\,e^{-2|\alpha|^2}$. 

Similar arguments apply if we consider the 
eigenbasis of $\rho_1$ which is related to 
the eigenbasis of $\rho_0$  by a unitary matrix $U$. 
The transformed matrix  
$(\mathbbm{1}_A\otimes U)\rho_{AB} (\mathbbm{1}_A\otimes U^\dag)$ 
must be of a form  similar to (\ref{eq:rho_0_AB}). Then, 
by comparison of the off-diagonal blocks one obtains the 
equality 
\begin{equation}
\label{eq:CUContra}
\begin{bmatrix}
 1  & 0 \\
     0  & 0 
  \end{bmatrix} =
 U
 \begin{bmatrix} 
 1  & 0 \\
     0  & 0
  \end{bmatrix} U^\dag,
\end{equation}
which implies that $U$ is diagonal in the chosen basis 
with some complex entries on the diagonal of modulus one. From this it follows that 
$|\phi_1\rangle = e^{i\lambda}|\phi_0\rangle, \, 
\lambda \in \mathbbm{R}$, which was excluded in the 
beginning.
So we have derived a contradiction to the assumption 
of separability simply by making use of the purity 
of the states measured at B and the knowledge of the 
reduced density matrix $\rho_A$.

In the following, we will extend this idea to the case
of outcome states which are affected by noise
\footnote{A first scheme for this task has been given in 
\cite{rigas05a}, though the tools developed in the present 
article allow a more systematical investigation.}. 
To this aim, 
we first need an adequate and powerful description of
quantum states of a qubit and a mode. In the next Section
we will introduce the so-called expectation value matrix 
for this task. 
Then, we will formulate separability criteria in this 
description. With these criteria, we can then investigate
the presence of entanglement in the actual QKD protocol.

\section{Classification of the Expectation Value Matrix}
\label{sec:EVMclas}

In our protocol, the density matrix has a special structure. 
While Alice has a discrete system, Bob's system consists of
a continuous variable system, namely a mode.

On the one hand, there exist efficient operational 
entanglement criteria for bipartite discrete systems 
considering the system's 
density matrix \cite{peres96a,horodecki96a,horodecki99a, 
rudolph03a, chen03a}. On the other hand, criteria for bipartite CV systems
exploiting uncertainty relations \cite{duan00a} and covariance 
matrices \cite{simon99a,werner00a,giedke01b} of quadrature operators 
measured on the whole state are known. 

One might be tempted to employ these CV entanglement criteria
for our half-discrete, half-continuous problem (e.g.~by
describing Alice's  discrete subsystem  in terms of two Fock 
states with different photon numbers).  However, it has been 
shown in Ref.~\cite{rigas05a} that  these criteria can not be 
successfully applied here, due to the limited knowledge on the 
whole bipartite state in our PM scheme.

Therefore, we introduce in this section a  quantity that 
describes the two different systems in their standard ways
and includes all properties accessible in a PM scheme using
heterodyne detection. Additionally, the basic properties of 
this object are derived.

\subsection{Definition}
\label{sec:EVMdef}
We introduce the  
\emph{bipartite expectation value matrix} (EVM) $\chi 
\in \mathbbm{C}^{6 \times 6}$ as 
\begin{equation}
  \label{eq:EVMdef}
  \chi := 
  \begin{bmatrix}
    \Big\langle |0\rangle\langle 0|\otimes B \Big\rangle & 
   \Big\langle |0\rangle\langle 1|\otimes B \Big\rangle \\
    \Big\langle |1\rangle\langle 0|\otimes B \Big\rangle & 
 \Big\langle |1\rangle\langle 1|\otimes B \Big\rangle  
  \end{bmatrix}
\end{equation}
with $B$ being the operator-valued matrix
\begin{equation}
  \label{eq:OpMatrixB}
  B:=
  \begin{bmatrix}
    \mathbbm{1} & x & y\\
    x & x^2 & \mathcal{S} (xy)\\
   y & \mathcal{S} (xy) & y^2
  \end{bmatrix}.
\end{equation}
In this definition, $x$ and $y$ denote the quadrature 
operators, obeying the commutation relations 
$[x,y] = xy-yx = i.$ Furthermore, $\mathcal{S}(xy)$ 
denotes the symmetrized product $(xy+yx)/2$
and $\langle A\otimes B \rangle=Tr(\rho A \otimes B)$ 
denotes the matrix of expectation values of the tensor 
product of $A$ with all operators of $B$ in a given state 
$\rho_{AB}$.

We take Alice's natural basis $\{|0\rangle,|1\rangle\}$ so 
that along with the identity on  Bob's side,  the elements 
of the reduced  density matrix $\rho_A$ are included (see 
next section). The two projectors $|0\rangle\langle 0|$ 
and $|1\rangle\langle 1|$ ensure that conditional
quadrature  expectation values and moments easily accessible 
to Bob are involved. 

Explicitly, with the  definitions (\ref{eq:EVMdef}) and
(\ref{eq:OpMatrixB})  and the specific form of $\rho_{AB}$ 
(\ref{eq:rho_AB}), the upper 
left $3\times 3$-block in 
(\ref{eq:EVMdef}) becomes
\begin{equation}
  \label{eq:Eta0def}
    \Big\langle |0\rangle\langle 0|\otimes B \Big\rangle = p_0
  \begin{bmatrix}
    1 & \langle x\rangle_0 & \langle y\rangle_0\\
\langle x\rangle_0 & \langle x^2\rangle_0 & \langle \mathcal{S} (xy) 
\rangle_0\\  
\langle y\rangle_0 & \langle \mathcal{S}(xy) \rangle_0 & \langle
y^2\rangle_0   
  \end{bmatrix} =: p_0 \,\eta_0,
\end{equation}
with the expectation value $\langle b\rangle_0:=
\mbox{Tr}(b\rho_0)$,
defining $\eta_0$ as the \emph{single mode's} EVM of the state
$\rho_0$. Obviously, 
$\eta_0$ is real and  symmetric by
construction.

\subsection{Properties}
\label{sec:EVMprop}
We now want to derive properties of the  EVM $\eta_0$ of $\rho_0$. 
For this, let us first look at the covariance 
matrix $\gamma_0$ of a single mode, defined as  
\begin{equation}
  \label{eq:covar}
  \gamma_0 := 
  \begin{bmatrix}
    \Delta(x)_0 & \Delta (\mathcal{S}(xy))_0\\
\Delta (\mathcal{S}(xy))_0 & \Delta(y)_0
  \end{bmatrix}
\end{equation}
with  $\Delta(x)_0 = \langle x^2 \rangle_0 - \langle x\rangle_0^2$ 
and $\Delta(\mathcal{S}(xy))_0 = \langle\mathcal{S}(xy)\rangle_0 -
\langle x\rangle_0\langle y\rangle_0$ etc.  
For such a matrix, it is known that a necessary and 
sufficient criterion of being a covariance matrix of a physical 
state is 
\begin{eqnarray}
\label{eq:Gamma2} 
  \gamma_0 +\frac{i}{2}J & \geq  & 0 \mbox{ with } J= 
  \begin{bmatrix}
    0 & -1\\
    1 & 0
  \end{bmatrix}.
\end{eqnarray}
This condition is an implementation of the canonical
commutation relations and the resulting uncertainty relations
obeyed by $x$ and $y$ \cite{robertson34a,giedke01a}. Note 
that complex 
conjugation of Eq.~(\ref{eq:Gamma2}) yields 
$\gamma_0 -\frac{i}{2}J  \geq   0,$ hence 
$\gamma_0 \geq   0$ has to hold, too.
 
Now we will show that for $\eta_0$ a condition similar 
to Eq.~(\ref{eq:Gamma2}) holds, namely 
\begin{equation}
\label{eq:EtaCond1}
\left( \eta_0 + \frac{i}{2}\tilde J\right) \geq 0 \mbox{ with }
 \tilde J :=  
  \begin{bmatrix}
    0 & 0 & 0\\
    0 & 0 & -1\\
    0 & 1 & 0
  \end{bmatrix}\,.
\end{equation}
Indeed, using the commutation relations $[x,y]=i$ one can 
see that
\be
\eta_0 + \frac{i}{2}\tilde J =
\begin{bmatrix}
\mean{\eins}_0 & \mean{x}_0 & \mean{y}_0\\
\mean{x}_0 & \mean{x^2}_0 & \mean{yx}_0\\ 
\mean{y}_0 & \mean{xy}_0 & \mean{y^2}_0
\end{bmatrix}\,,
\ee
which is clearly positive since any term 
$\vec{s}^\dagger (\eta_0 + i \tilde J /2 ) \vec{s}$ 
with $\vec{s}=(s_1,s_2,s_3)^T$ can be expressed as 
$\mean{A^\dagger A}$ with $A=s_1 + s_2 x +s_3 y$. 

Obviously, the positivity condition (\ref{eq:EtaCond1}) does 
not depend on a particular measurement on Alice's side but must 
hold for any projector
$|s\rangle\langle s|$ evaluated by A, i.e., for any $\rho_s:=
\mbox{Tr}_A(\rho_{AB}|s\rangle\langle s| \otimes \eins)$, the corresponding 
EVM $\eta_s$ must  fulfill 
$(\eta_s + {i}\tilde{J}/2) \geq 0.$

Furthermore, from Eqs.~(\ref{eq:EVMdef}) and (\ref{eq:OpMatrixB}) 
it is obvious that for $i,j =1,4$, the sub-matrix $[\chi]_{ij}$ 
is exactly the reduced density matrix $\rho_A$, since on  Bob's 
side, only the  identity is  evaluated. By construction, we have 
also $\chi=\chi^\dag.$
So we can summarize:

\begin{observation}
  \label{obs:ChiCond}
For any bipartite state $\rho$, its EVM $\chi(\rho)$ (as defined in
Eq.(\ref{eq:EVMdef})) has the following properties:
\begin{itemize}
\item $\chi$ is Hermitean: $\chi \,=\, \chi^\dag.$
\item for $i,j=1,4$, the sub-matrix $[\chi]_{ij}$ is the reduced
  density matrix $\rho_A$ (and thus, positive, Hermitean and of 
unit  trace) 
\item for any projector $|s\rangle\langle s|$, the  EVM $\eta_s:=
  Tr_A\left\{(|s\rangle\langle s|\otimes 
    \mathbbm{1}_B)\chi\right\}$ of the corresponding
  mode must satisfy
  \begin{equation}
    \label{eq:EtaS}
\left( \eta_s + \frac{i}{2}\tilde
  J\right) \geq 0 .
  \end{equation}
This condition implies $(\eta_s - i \tilde{J}/2) \geq 0$
and $\eta_s \geq 0$ as well.
\end{itemize}
\end{observation}

\subsection{Separability conditions}
\label{sec:SepCond}

Now we want to derive necessary conditions for separability
in terms of the EVM. We will start with pure product states.

For a pure product state 
$\rho= |s\rangle\langle s| \otimes \rho_s$, the bipartite EVM
$\chi(\rho)$ is of the form $|s\rangle\langle s| \otimes
\eta_s$. Since the projector $|s\rangle\langle s|$ is positive, by
virtue of Eq.(\ref{eq:EtaS}) $\chi$
must then fulfill 
\begin{equation}
\label{eq:ProdChi}
 \chi \pm\left( |s\rangle\langle s| \otimes \frac{i}{2}\tilde{J}\right)
=\ketbra{s}\otimes \left( \eta_s \pm \frac{i}{2}\tilde{J}\right)
\geq 0 \, .
\end{equation}
For a  general separable state, which can be written as  
$\rho=\sum_k p_k \rho_k^A \otimes \rho_k^B$, the EVM 
$\chi(\rho)=\sum_k p_k \rho_k^A\otimes\eta_k^B$
must hence satisfy
\begin{equation}
\label{eq:ChiCond}
\chi \pm \left(\frac{i}{2}\rho_A\otimes\tilde J\right)\geq 0 \, .
\end{equation}
The second relation holds due to Eq.~(\ref{eq:ProdChi}) and 
the fact that $\sum_k p_k \rho_k^A = \rho_A$.
We  can summarize:
\begin{observation}
  For any separable state $\rho$,  its EVM $\chi$ additionally to the
  conditions specified in Observation \ref{obs:ChiCond} must satisfy
  the   following inequalities:
 \begin{equation}  \label{eq:MainCond}
    \chi \pm \frac{i}{2}\rho_A\otimes \tilde J \geq  0.
  \end{equation}
  Note that this inequality implies $\chi\geq 0$, as well.
\end{observation}

\section{Application to a general PM Scheme}
\label{sec:Application}

Let us now  connect these necessary  conditions on the 
EVM of a bipartite separable state $\rho$ to the knowledge 
on that state accessible in any PM scheme with two signal 
states and heterodyne detection. Given the available
entries of the EVM we derive a set of matrix inequalities
which have to be fulfilled. The question whether 
they can be fulfilled can then be solved efficiently by 
semidefinite programming.

\subsection{Knowledge on $\chi$ in PM schemes}
\label{sec:Knowledge}
Let us first determine the entries in the EVM which are
accessible in any PM scheme. For  $|s\rangle_A =
|0\rangle_A$ or $|1\rangle_A$, $\eta_s$ corresponds 
to Bob's measurement outcomes under the condition that 
A  sent signal 0 or 1. Bob has the full information on 
these states $\rho_0$ and $\rho_1$, i.e., all expectation 
values in $\eta_0$ and $\eta_1$ are  fully known. With  
knowledge of the a priori probabilities $p_0,\,p_1$ this 
gives full information on $\chi_{ij}$ for $ i ,j = 1,2,3$ 
or $i,j =4,5,6$, i.e. for the upper left and lower
right $3\times 3$-block of $\chi$ (c.f. Eq.(\ref{eq:EVMdef})). 

For $i= 1,2,3 \,, j =4,5,6 $ and vice versa, the only
operator product of $|0\rangle\langle 1|\otimes B$ in 
Eq.(\ref{eq:EVMdef}) that can be evaluated is 
$|0\rangle\langle 1|\otimes \mathbbm{1}$ (or 
$|1\rangle\langle 0|\otimes \mathbbm{1}$, respectively;
c.f. Eq.(\ref{eq:OpMatrixB})) because they are known from 
the reduced density matrix of Alice. 

It is important to note at this point that since 
dim$(\mathscr{H}_A)=2$, we can (and will) always 
choose $\rho_A$ to be real and thus have 
$\rho_A =\rho_A^T$ by a proper phase choice of 
$|0\rangle$ and $|1\rangle$. Obviously, this
property holds for arbitrary signal states sent to Bob.   

The remaining 16 entries  $\chi_{ij}$ (for $i=1,2,3
\,,j =4,5,6$ and vice versa with $\{i,j\}\neq\{1,4\}$) are 
unknown but can be further restricted by the conditions 
in Observation \ref{obs:ChiCond}  to five free complex 
parameters.

The explicit form of $\chi$ then becomes
\begin{equation}
  \label{eq:ChiExplicit}
  \chi = 
  \begin{bmatrix}
    p_0 
\eta_0 &
    \begin{array}{ccc}
   S & a & b\\
   a & c & d\\
   b & d & e
    \end{array} \\
    \begin{array}{ccc}
  S & a^* & b^*\\
  a^* & c^* & d^* \\
  b^* & d^* & e^*
  \end{array}&p_1
\eta_1
  \end{bmatrix}
\end{equation}
with $S$ abbreviating $(\rho_A)_{01}$ and  $a,b,c,d,e
\in \mathbbm{C}$ being free complex parameters. From 
Observation 1,  we have the following general result
holding for all protocols with two signal states and 
heterodyne detection:
\begin{observation}
\label{obs:Theorem1}
Let $\rho_A, \eta_0~and~\eta_1$ be specified by a certain 
set of measurement data in  an arbitrary PM scheme with 
two signal states and heterodyne detection. If no set of 
parameters $a,b,c,d,e \in \mathbbm{C}$ can be found  
such that $\chi$ as specified in Eq.(\ref{eq:ChiExplicit}) 
satisfies the inequalities (\ref{eq:MainCond}), then the 
measured bipartite state $\rho$ must have been entangled. 
Consequently, if such a set of parameters can be found, 
then the QKD protocol is insecure.
\end{observation}

It is possible to show that one can concentrate on real 
parameters $a,b,c,d,e$, only: 
\begin{lemma}  
\label{lem:Real}
Suppose $X=\chi$ is a solution of the form (\ref{eq:ChiExplicit}) to a
problem specified  by data $\eta_0, \eta_1,\rho_A$. Then there 
exists always a real solution $\ov{X}\in \mathbbm{R}^{6\times 6}$.
\end{lemma}

Proof: Let us first show that $X^T$ is also a solution. Since
$\eta_0^T=\eta_0$ and $\eta_1^T=\eta_1,$ $X^T$ still fits to 
the experimental parameters. Furthermore, since we have chosen
$\rho_A=\rho_A^T,$ the new $X^T$ obeys still the inequalities
(\ref{eq:MainCond}), since $\tilde{J}^T=-\tilde{J}.$ 
But then $\ov{X} :=(X + X^T)/2$ is another solution, which
is real.
$\qed$

Finally we show that with all knowledge available in a PM scheme, 
PPT entangled states cannot be distinguished from separable states: 

\begin{lemma}
\label{lem:PPT}
Suppose a PPT-entangled state $\rho$  compatible with data
$\eta_0,\eta_1,\rho_A$, i.e. the EVM $\chi(\rho)$ is a solution to a
problem as specified in Observation \ref{obs:Theorem1}.  
Then there exists a separable state $\ov{\rho}$ which is also
compatible with the data, i.e. whose EVM is also a solution to the
same problem. 
\end{lemma}

Proof: Since $\rho$ is PPT, $\rho^{T_A}$ is a valid physical 
state, too. From the construction of the EVM (\ref{eq:EVMdef}) 
it can be seen immediately that the EVM of $\rho^{T_A}$ equals  
$\chi^{T_A}(\rho)$, i.e., the partial transpose of the EVM 
$\chi$ of $\rho$. Furthermore, we have $\chi^{T_A}=\chi^{T}.$
So for 
$$
\ov{\rho} := \frac{1}{2}(\rho + \rho^{T_A})
$$
its EVM $\ov{\chi}=(\chi + \chi^{T_A})/2 = 
(\chi + \chi^{T})/2$ is a solution to the
problem specified by the data as shown in the proof of Lemma
\ref{lem:Real}. Since $\ov{\rho} = \ov{\rho}^{T_A}$ and 
dim$\mathscr{H}_A = 2$, $\ov{\rho}$ is separable (Theorem 2 in
\cite{kraus00a}). So the PPT-entangled state $\rho$ and the 
separable state 
$\ov{\rho}$ are both compatible with the available data.
$\qed$ 

\subsection{Implementation with semi-definite programming}
\label{sec:SDP}

It is hard to find analytically a set of  
parameters $a,b,c,d,e$ to a given matrix $\chi$ 
(\ref{eq:ChiExplicit}) with $\rho_A,\eta_0,\eta_1$ 
fixed so that Ineqs.(\ref{eq:MainCond}) are indeed 
fulfilled. However, this task can be easily implemented 
and efficiently solved with semi-definite programming 
\cite{vandenberghe96a}. 

A semidefinite program is a convex optimization problem
of the type
\begin{equation}
\label{eq:MinSDP}
\min_x c^T x
\end{equation}
subject to
\begin{equation}
  \label{eq:SDPconstr}
  F_0 + \sum_{k=1}^N x_k F_k \geq 0.
\end{equation}
Here, $x \in \mathbbm{R}^N,$ $c \in \mathbbm{R}^N,$  
and the $F_0,\,F_k, \,k=1\ldots N$ are Hermitean matrices.
The matrix inequality (\ref{eq:SDPconstr}) defines a convex
subset in the vector space $\mathbbm{R}^N$.

Optimization problems of this type have several nice 
properties \cite{vandenberghe96a}. 
For usual minimization problems it is 
impossible to guarantee that an obtained solution 
is really the {\it global} minimum. This is not the 
case for semidefinite programs, since here the 
so-called dual problem delivers a lower bound on the 
minimum. Under weak conditions, this lower bound coincides 
with the minimum, thus global optimality of a solution may 
be proved. Furthermore, efficient algorithms for the 
implementation of semidefinite programs are freely available
\cite{sedumi,sdpt,yalmip}.

To implement the constraints in Observation \ref{obs:Theorem1}, 
the question of interest is whether or not there exists a
solution of the form (\ref{eq:ChiExplicit}) to given data 
matrices $\rho_A, \eta_0, \eta_1$ which satisfies the two 
inequalities (\ref{eq:MainCond}). In the language of 
semidefinite programming it is only of interest whether 
the constraints in  (\ref{eq:SDPconstr}) can be fulfilled. 
This is a so-called feasibility problem, where the objective 
function  (\ref{eq:MinSDP}) can be ignored.

The data matrices   $\rho_A, \eta_0,\eta_1$ are included in 
$F_0$ as well as $\pm (i/2)\rho_A\otimes \tilde J$, while the 
free real parameters $a,b,c,d,e$ form the vector 
$x\in\mathbbm{R}^5$. The specific form (\ref{eq:ChiExplicit}) 
of $\chi$ then determines the shape of the real symmetric matrices 
$F_k, \, k=1\ldots 5$. If the problem is returned infeasible, then 
the bipartite state  must have been entangled. 

\begin{figure}
\centering
\includegraphics[scale=0.5]{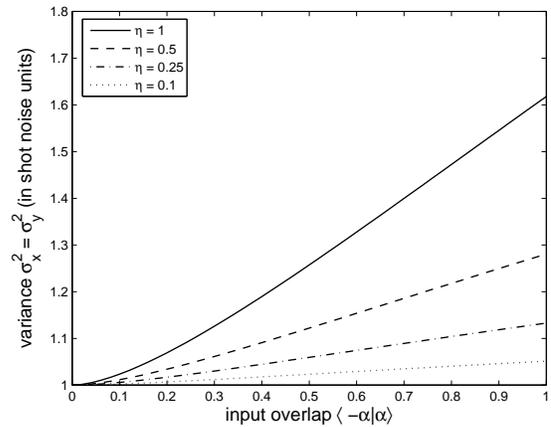}
\caption{Entanglement detection with semi-definite programming. For each value 
  of 
  transmission, the area below the corresponding line belongs to
  parameter pairs outcome overlap vs. quadrature variance for which
  entanglement can be ensured.}
  \label{regime}
\end{figure}

In order to illustrate our method, let us choose coherent 
states $\ket{\pm \alpha}$ as input states. Then, we set the
a priori probabilities $p_0, p_1$ both to $1/2$ thus 
$S=(\rho_A)_{01}$ in Eq.~(\ref{eq:ChiExplicit})
becomes $\exp(-2|\alpha|^2)/2$.
Now, let us assume  measurement 
outcomes of $\rho_{0/1}$ at Bob's site generating EVMs
\begin{equation}
  \label{eq:Eta01}
  \eta_{0/1} = 
  \begin{bmatrix}
    1 & \pm  c & 0 \\
 \pm   c & c^2 + \sigma^2 & 0\\
   0 & 0 & \sigma^2
  \end{bmatrix}
\end{equation}
with symmetric quadrature variances $\sigma_x^2=\sigma_y^2$.
The quadrature expectation values are set to 
$\langle x\rangle_{0/1} =\pm c$ and $\langle y\rangle_{0/1}=0$. 
We also assume  vanishing expectation values for $\mathcal{S}(xy)$. 
Note that this property does not necessarily mean that the outcome 
states $\rho_0, \rho_1$ are Gaussian.  
The performance is shown in Fig. \ref{regime} for several 
transmission values $\eta := c^2/|\alpha|^2$.

All lines show a similar behavior: For an input overlap
$\langle-\alpha|\alpha\rangle $ close to zero, the outcome states
are quite distinguishable, too. In this case, the outcomes must be
extremely pure in 
order to show entanglement. The more the input states are
overlapping, the more noise can be tolerated for a certain amount of
loss. 

For an overlap of 1 (i.e., $\alpha=0$), however, the input
states factorize off from Alice's logical qubits, so they are no longer
entangled. Thus, in the limit of the overlap going to 1, the graph
drops down to 
$\sigma^2=1$ discontinuously.

It has to be emphasized that even for 90\% loss, the ability of
detecting entanglement is still well within a reasonable 
tolerance of noise, achievable in current experiments.
Also, for all transmission values $\eta$, the necessary relation
between transmission and excess noise $\delta := \sigma^2 -1$, given
by $\delta < 2\eta$ \cite{namiki04a}, is satisfied.

\subsection{Generalizations}
So far we considered a scheme where we send two coherent states and perform a 
heterodyne measurement to extract information about the covariance matrix and 
the expectation values of two quadrature operators. Any quantum mechanical 
measurement which allows us to infer these observable quantities will suffice 
to proof the presence of entanglement with our method.

One can also consider a situation where only $\mean{x^2}$ and $\mean{y^2}$ are 
measured but not $S(x y)$. This is the case when Bob measures only two conjugate 
quadratures by homodyne measurements. This leads to an additional free parameter 
$f$ that replaces $S(x y)=0$ in Eqn.(\ref{eq:Eta01}). By numerical evaluation we 
found that the parameter regime,  shown to be incompatible with separable states 
by our approach, is exactly the same as if we had measured $S(x y)=0$.

Note that our analysis does not make use of the explicit form of the signal states. 
Instead of coherent states one could have used any two quantum mechanical states. 
Only the overlap between the two states is relevant and enters the analysis. 

\section{conclusions}
\label{sec:conc}
In conclusion, we have investigated separability properties 
of quantum states consisting of a qubit and a mode. We 
introduced the EVM matrix as a suitable description of 
such systems and derived a necessary separability 
criterion in this formulation. For reduced information, 
this separability criterion can be efficiently checked via
semidefinite programming. We have then applied these results 
to a general PM QKD protocol using coherent signal states and 
heterodyne detection.  Also an extension to homodyne measurement of two 
conjugated variables only has been given. We showed that PPT 
entanglement cannot be 
detected in this scheme. For realistic setups, however, we 
calculated that entanglement detection is possible even
in the case of high transmission losses.

\begin{acknowledgments} 
We thank Stefan Lorenz, Tobias Moroder and Volkher Scholz 
for valuable discussions. 
This work has been supported by the DFG (Emmy Noether Programm)
and by the EU (OLAQUI, PROSECCO, QUPRODIS, SCALA, SECOQC)
and the FWF.

\end{acknowledgments} 

%\bibliographystyle{apsrev}
%\bibliography{librarium}

\end{document}